\documentclass[final,12pt]{article}

\usepackage{graphicx}
\usepackage{amssymb}
\usepackage{authblk}
\usepackage{hyperref}

\def\MeV{\mbox{MeV/c$^2$}}
\def\GeV{\mbox{GeV/c$^2$}}
\def\ra{\mbox{$\rightarrow$}}

\def\ttbar{\mbox{$t\overline{t}$}}

\def\pp{\mbox{$p\overline{p}$}}
\def\qq{\mbox{$q\overline{q}$}}
\def\ud{\mbox{$u\overline{d}$}}
\def\du{\mbox{$d\overline{u}$}}
\def\cs{\mbox{$c\overline{s}$}}
\def\sc{\mbox{$s\overline{c}$}}
\def\us{\mbox{$u\overline{s}$}}
\def\sd{\mbox{$s\overline{d}$}}

\title{Mass biases in reconstructing exclusive radiative hadronic decays of W bosons at the LHC.}
\author[1]{W. J. Murray}
\affil[1]{University of Warwick \&  STFC RAL}
\author[2]{ E. Jones}
\affil[2]{University of Warwick}

\date{}

\begin{document}

\maketitle

\begin{abstract}
\noindent The search for exclusive hadronic vector boson decays is an ongoing part of the LHC programme where, to date, no such decays have been observed. In addition to the intrinsic interest in the branching ratios, there is potential for a measurement of the $W$ boson mass quite distinct from the usual methods. The radiative decay modes offer good potential channels for this search; however, we highlight three issues with it not previously discussed: particle misidentification, partial reconstruction and the impact of interference with QCD. These issues cause shifts in the peak position of tens or hundreds of \MeV.
\end{abstract}

\section{Introduction}
\label{sec:intro}
Exclusive decays of  $W$ and $Z$ bosons to hadronic final states have never been observed. Low-multiplicity final states could potentially be experimentally accessible and, in particular, radiative decays of vector bosons are promising channels with many theoretical predictions of the branching ratios available~\cite{AMP,GKN,HP,MM14,KP}. However, there is not universal agreement between these predictions. For example, the $W^{\pm}\ra\pi^{\pm}\gamma$ branching ratio is estimated to be of the order of $10^{-9}$ in Ref.~\cite{AMP} and Ref.~\cite{MM14}, whereas Ref.~\cite{KP} proposes values as high as $10^{-7}$.

There are experimental limits on many two-body $Z$ decays. These include radiative decays to final states such as $\pi^0\gamma$~\cite{CDF14}, $\eta\gamma$ and $\eta^\prime\gamma$~\cite{ALEPH92}, $\omega\gamma$~\cite{DELPHI94}, $\phi\gamma$~\cite{ATLAS16}, $J/\psi\gamma$ and $\Upsilon\gamma$~\cite{ATLAS15}, as well as non-radiative modes such as $\pi^0\pi^0$~\cite{CDF14}. There are fewer experimental limits on similar $W$ decays, although limits on $W^-\ra\pi^-\gamma$ at $1.51\times 10^{-5}$ have been presented recently by CMS~\cite{CMS21}. The analysis uses $W$ bosons produced in \ttbar~events, with a leptonically-decaying $W$ boson  used to trigger and reduce backgrounds, whilst the other $W$ boson is used for the decay search. The use of this trigger strategy comes at the cost of a lower rate, but given that the total LHC $W$ cross-section is much larger, approximately 175~nb at 13~TeV according to DYNNLO 1.5~\cite{CG07,C09}, over $5\times 10^{11}$ $W$ bosons are expected at the HL-LHC. Therefore, with a suitable trigger, this presents an interesting opportunity, and might allow for a $W$ mass measurement to be possible, a measurement previously discarded on grounds of rate in $t\overline{t}$ by Ref.~\cite{MM14}. 

The current world average mass of the $W$ boson is $80.379\pm0.012$ \GeV~\cite{PDG}. This is calculated from a combination of results using $p\bar{p}$, $e^+e^-$ and $pp$ interactions at a variety of experiments. There are ongoing studies at the LHC in $pp$ interactions, where the $W$ boson mass is computed from the reconstructed transverse mass using the leptonic decay modes. Precision measurements of the properties of the $W$ boson, including its mass, are essential in testing the predictions of the Standard Model. 

ATLAS has demonstrated a dedicated trigger for two body radiative $Z/H$ decays~\cite{ATLAS16,Aaboud:2017xnb}. It was used in 2015 to select $\phi\gamma$ candidates and was extended in 2016 to include $\rho\gamma$ candidates. It required a 35~GeV $E_{T}$ photon and a pair of charged tracks consistent with the mass hypothesis studied, giving an efficiency at or above  75\%. We assume that a similar strategy could be employed for low-multiplicity, radiative $W$ boson decay modes. It may also be possible to record exclusive fully-hadronic events using either a $J/\psi$ decaying to muons or a modified tau-pair trigger at first level, with a tight selection on the mass in charged particles at high level, but we focus on the radiative decay topology here. 

The ATLAS and CMS detectors at the LHC achieve excellent mass resolution for decays to charged particles and isolated photons, with 1-2~\GeV\ quoted by CMS for $H\ra\gamma\gamma$ and $H\ra ZZ$~\cite{CMS12}. A nominal 1\% resolution is used in this paper for decays to such particles. This resolution, combined with the possibility of millions of signal events, suggests a measurement of the $W$ boson mass could be feasible if there are exclusive decay modes with a branching ratio close to the CMS limit above. This paper highlights previously undiscussed issues due to particle (mis)identification, partial reconstruction and EW-QCD interference, which would complicate a mass measurement and require an evaluation, even when simply searching for the decays. Unfortunately, the fragmentation models in Sherpa and PYTHIA are not capable of simulating the two body $W^{\pm}\ra h^{\pm}\gamma$ process, so the paper focusses on $W^{\pm} \ra (h^- h^+ h^{\pm}) \gamma$, the lowest-multiplicity radiative process where the isolated photon and the charged particle tracks can all be well measured in LHC detectors.

\section{Simulating radiative decays}
\label{sec:srd}

\begin{figure}[htb]
\centering
\includegraphics[width=12cm]{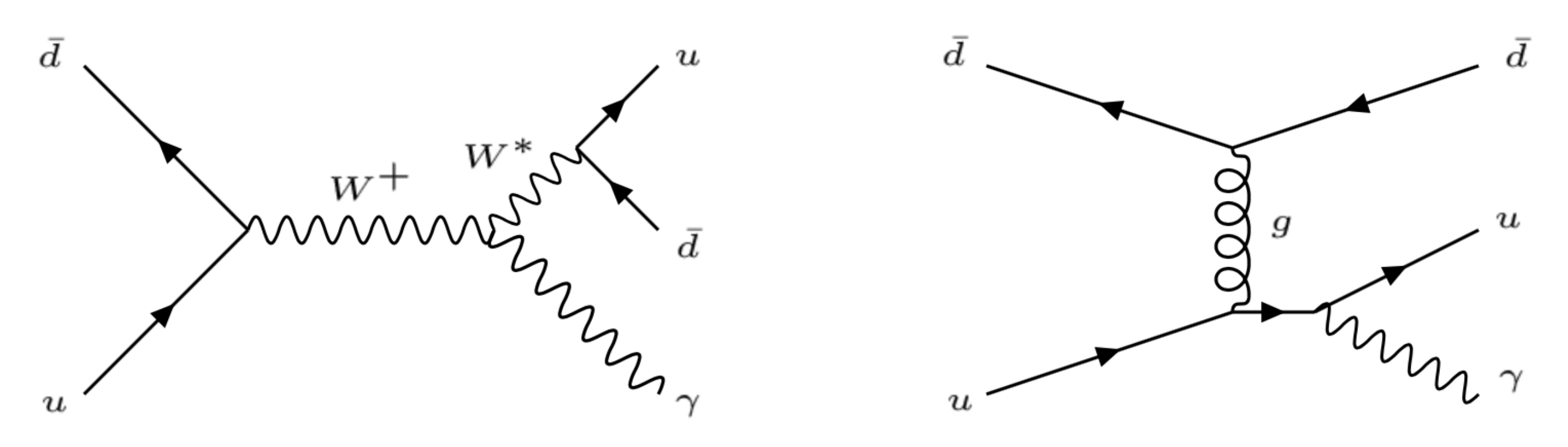}
\vskip -0.2cm
\caption[]{Feynman diagrams demonstrating two processes that have a \ud~pair recoiling against a photon in the final state. The left diagram shows the EW radiative process of interest, whilst the right shows an example QCD process that will lead to interference. \label{figs/Fig:feyn} }
\end{figure}
 
The signature studied here is that of a $W$ boson decaying to a photon recoiling against a low-multiplicity jet. An example Feynman diagram for this process is given in Fig.~\ref{figs/Fig:feyn}~(a). For most of the studies, we generate the inclusive process, $W^+\ra\ud$, and focus on the decays with a hard photon in the final state, comparing the decay multiplicity for PYTHIA8~\cite{PYTHIA6,PYTHIA8.1} and several different Sherpa 2.2.10 configurations~\cite{Sherpa,Sherpa-photon,comix}. The default Sherpa cluster-fragmentation model is not expected to be reliable~[F.Krauss, Personal communication] for processes with few particles in the final state, due to a mass of approximately 300~\MeV\ assigned to the quarks. Consequently, the Lund string fragmentation model is used for all alternative Sherpa configurations. This is done using the tune parameters suggested in the Sherpa manual: {\tt PARJ(21)=0.432; PARJ(41)=1.05; PARJ(42)=1.0; PARJ(47)=0.65; MSTJ(11)=5}. An alternate tune~\cite{Bothmann_2019}, produced no significant changes in the context of the work presented here. 

The Sherpa configurations tested are: replacing the cluster-fragmentation with the Lund string fragmentation (Sherpa-Lund); additionally enabling EW emission in the shower using the switch {\tt CSS\_EW\_MODE=1} (Sherpa-EW-Lund); and, lastly, using the hard matrix element to simulate $W^+\ra\ud\gamma$ (Sherpa-ME-Lund). This final configuration is done without the EW emission in the shower enabled to avoid double counting. The decays of $\pi^0$s and $K^0$s are disabled when counting particle multiplicity.

\begin{figure}[t]
\centering
\includegraphics[width=6.5cm]{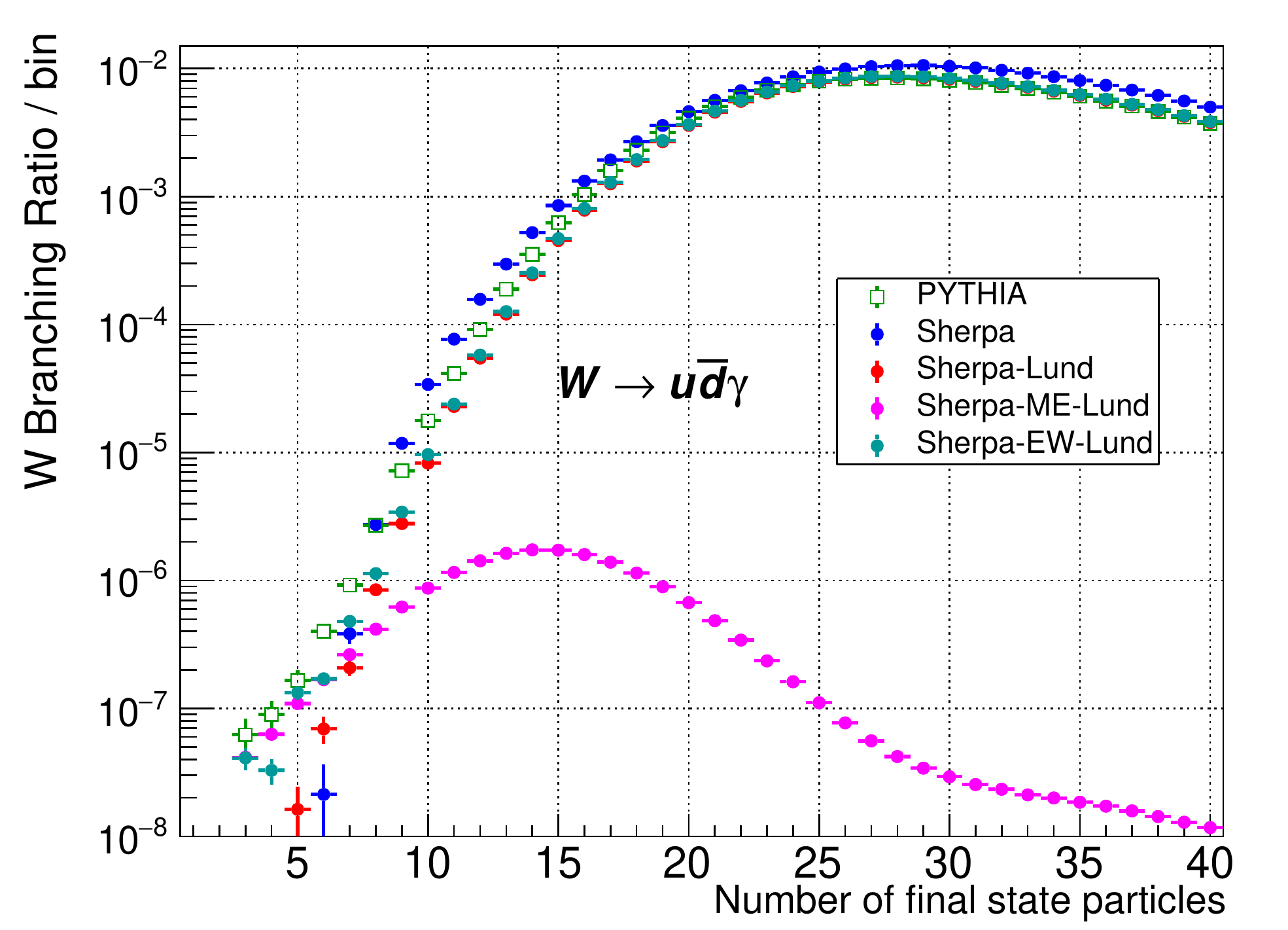}
\includegraphics[width=6.5cm]{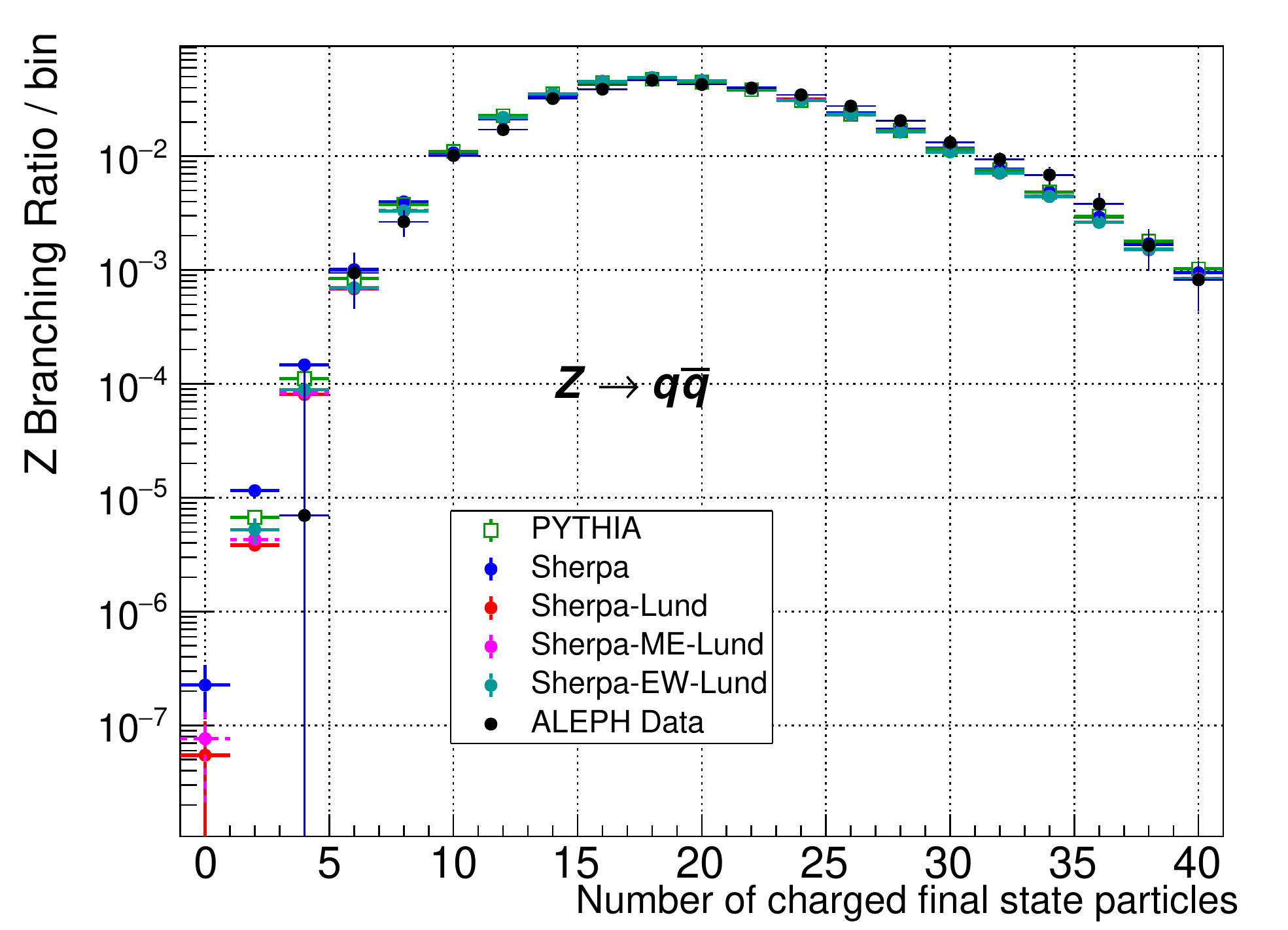}
\vskip -0.2cm
\caption[]{(a) shows the particle multiplicity for the process $W\ra\ud$, with at least one photon in the final state. (b) shows the charged-particle multiplicity for the process $Z \ra\qq$, with data from the ALEPH experiment overlaid. In both figures, PYTHIA is shown with open symbols and the predictions by the various Sherpa configurations are solid: default Sherpa (blue), Sherpa-Lund (red), Sherpa-EW-Lund (green) and Sherpa-ME-Lund (magenta). \label{figs/Fig:npart} }
\end{figure}

The distributions of the $W^+\ra\ud$ particle multiplicities produced by these generators when requiring at least one photon in the final state are shown in Fig.~\ref{figs/Fig:npart}~(a). The Sherpa-ME-Lund distribution required a photon energy above 30 GeV and the mass of the \ud\ system to be below 20~\GeV. Therefore, this setup can only be expected to be accurate in the extreme low-multiplicity regime. However, as the cross-section is four orders of magnitude lower in this regime, it can be much more efficiently simulated. In addition, with this configuration, both of the diagrams in Fig.~\ref{figs/Fig:feyn} can be produced coherently, allowing the study of interference effects, as discussed in Section 5. 

The default Sherpa cluster fragmentation has, overall, the highest branching ratio to at least one photon, which is mostly due to a higher production rate of $\eta$ mesons. At a multiplicity of ten, the predicted branching ratios (excluding the Sherpa-ME-Lund configuration) span a factor of five, with the largest differences being between the default Sherpa and the Sherpa-Lund configurations. However, at a multiplicity of 5 and below, PYTHIA, Sherpa-EW-Lund and Sherpa-ME-Lund, the only models producing hard photons, are within a factor of three of each other. This approximate agreement supports the use of the matrix-element approach for this extreme phase space. Sherpa-ME-Lund predicts a $W$ branching ratio to a radiative three-body state of $(4.1\pm0.2)\times 10^{-8}$, with the other simulations compatible within larger statistical errors. Two-body radiative states cannot be produced by the Lund hadronization model. 

In Fig.~\ref{figs/Fig:npart}~(b), we show the charged-particle multiplicity for the process $Z\ra\qq$ for the same configurations outlined above. Here the Sherpa-ME-Lund and Sherpa-Lund are summed, so that the negligible impact of including this radiative component is clear. In addition, we overlay data from the ALEPH experiment~\cite{ALEPH95}. The data does not disfavour any of the models and, at charged multiplicities of four or less, is not precise enough to be a definitive guide, though it hints that the true rate of extremely low-multiplicity events could be lower than these simulations predict. One interesting feature is that Sherpa, using its default cluster fragmentation, has the highest probability of producing fewer than five {\em charged} particles in $Z$ decay, while it has the lowest probability of giving fewer than five {\em total} particles in radiative $W$ decay. This remains true when the requirement for at least one photon is removed. This emphasises the uncertain nature of these distributions.

\section{Particle misidentification}
\label{sec:pm}

\begin{figure}[t]                  
\centering                        
\includegraphics[width=4.4cm]{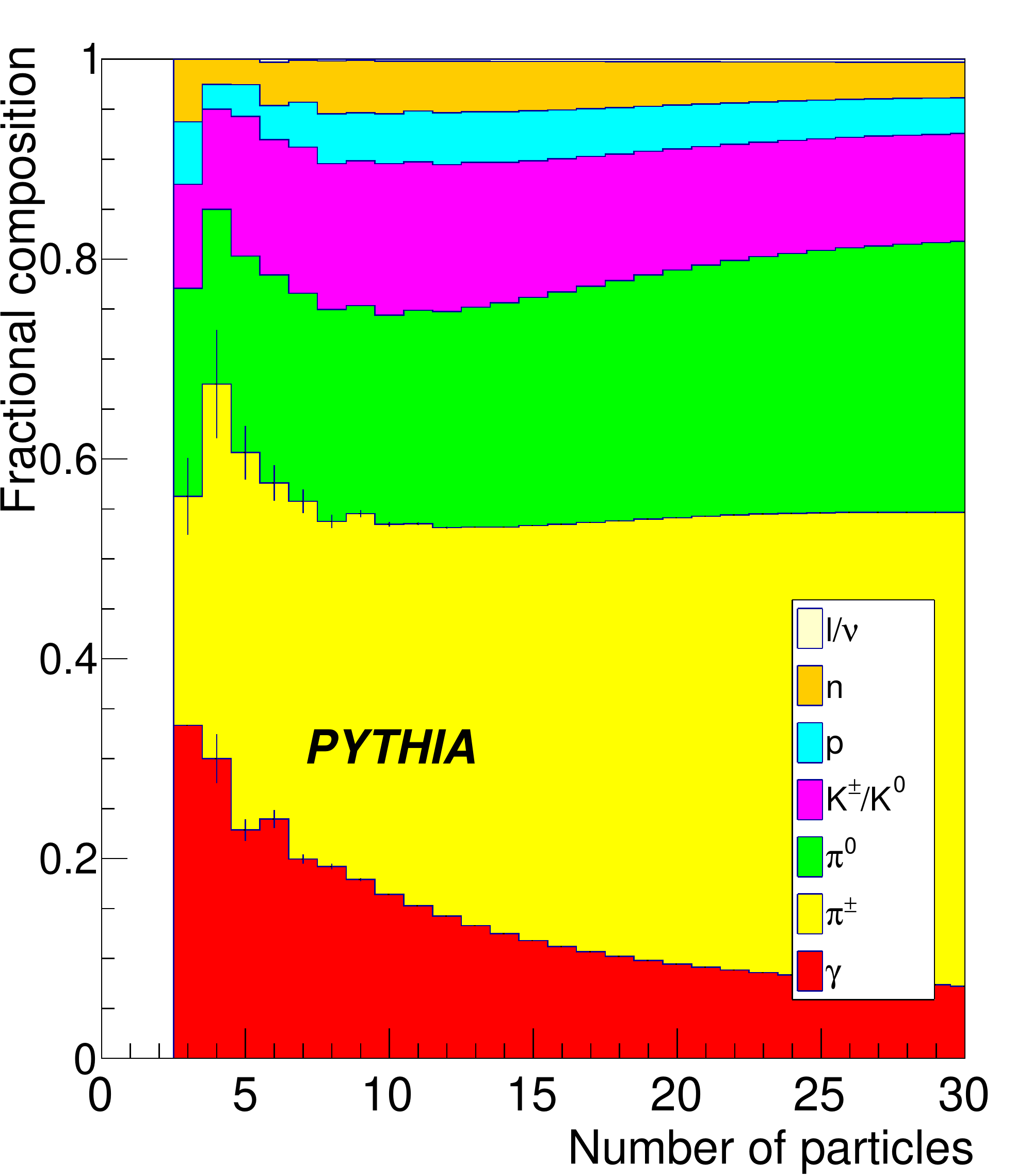}
\includegraphics[width=4.4cm]{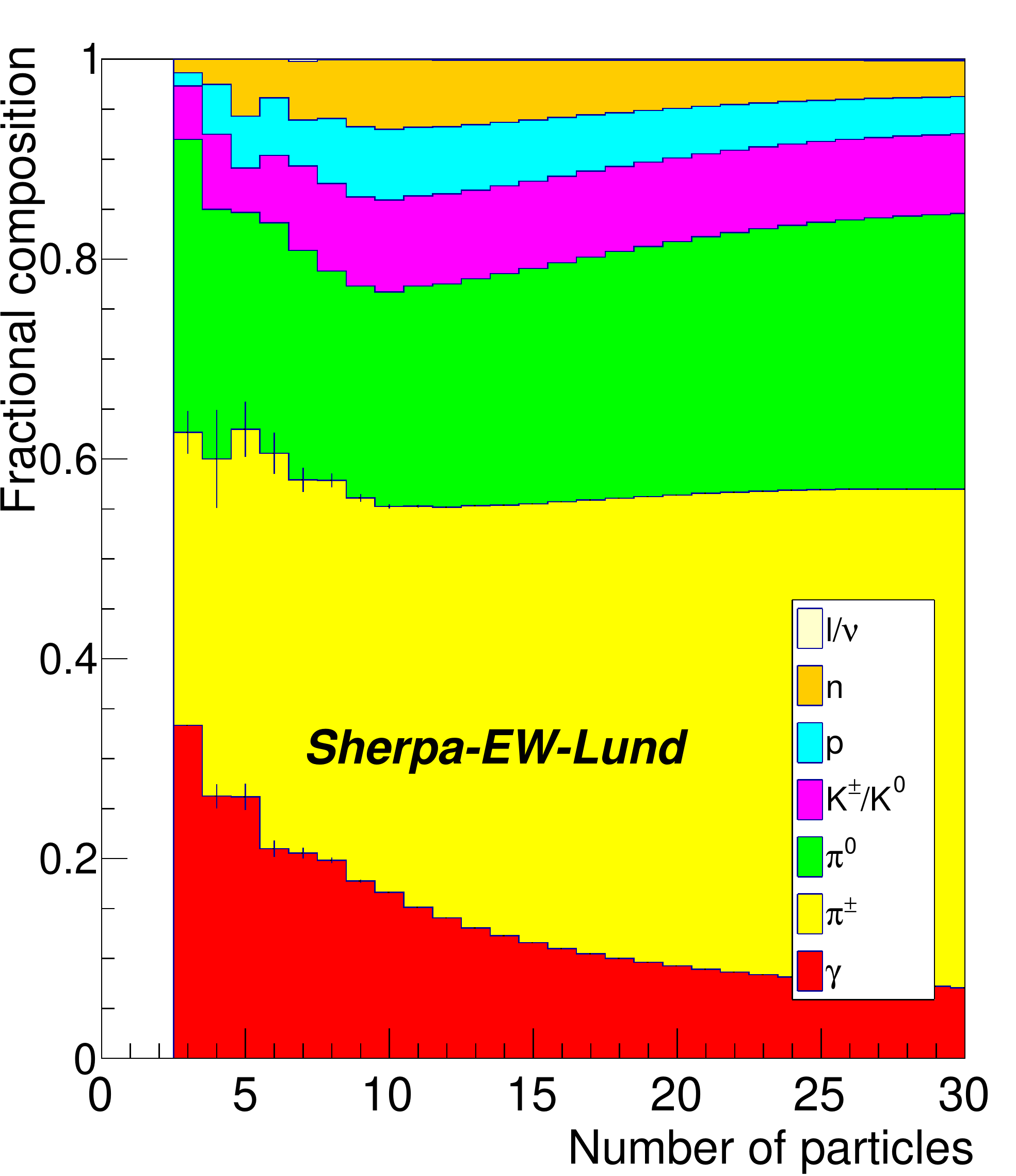}
\includegraphics[width=4.4cm]{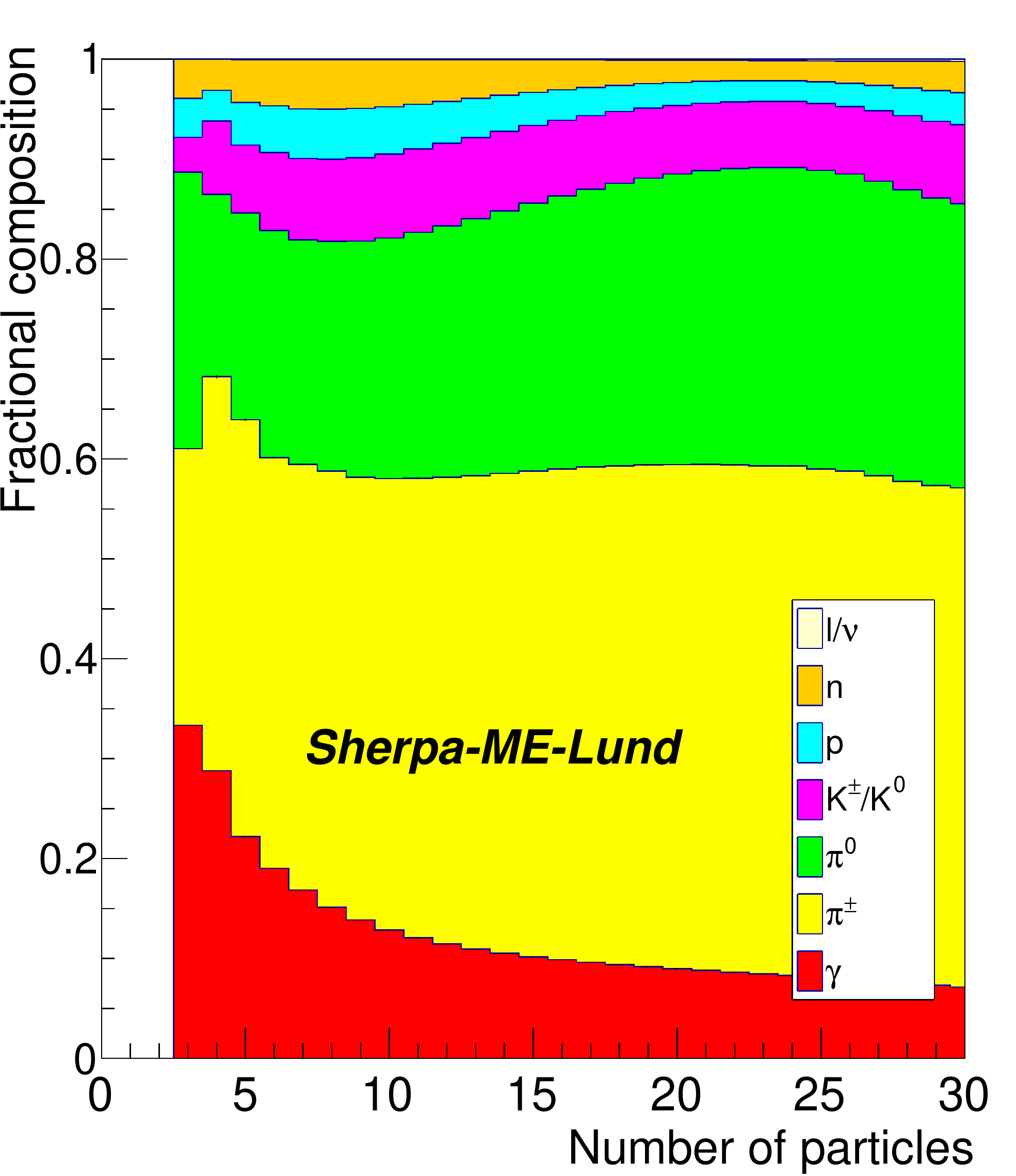}
\vskip -0.2cm                     
\caption[]{The fractions of leptons, neutrons, protons, kaons, $\pi^0$, $\pi^\pm$ and photons predicted by (a) PYTHIA, (b) Sherpa-EW-Lund and (c) Sherpa-ME-Lund for $W^+\ra\ud$ decays with at least one photon in the final state. Error bars at the top of the photon and charged pion components represent the error on that fraction alone; they are not shown on the other components to reduce clutter. \label{figs/Fig:compositiontotal} }
\end{figure}

The particle composition of low-multiplicity events predicted by the generators is shown in Fig.~\ref{figs/Fig:compositiontotal}, where at least one photon is required from all configurations. There are significant discrepancies in the low-multiplicity region where little data guidance is available. The most extreme example is the kaon fraction, which PYTHIA predicts to be roughly double that predicted by the other models. The statistical precision is much better in the Sherpa-ME-Lund sample, since it was generated with a low-mass hadronic system that allows for more efficient simulation; however, its predictions at high-multiplicity are not generally applicable.

Consequently, there is considerable uncertainty about the composition of low-multiplicity states and both ATLAS and CMS are not able to reliably distinguish hadron species at these momenta. If the wrong species of particle is attributed to a decay, the estimated boson mass will change. For example, if $W^+\ra\pp\pi^+\gamma$ is mistaken for $W^+\ra\pi^+\pi^-\pi^+\gamma$, the shift in the peak position of the $W$ boson mass is of the order 60~\MeV.

Both CMS and ATLAS have proposed plans to include timing detectors as part of the Phase II upgrades, primarily to target the challenges that will arise from increased pile-up at the HL-LHC~\cite{CMS:Timing,ATLAS:Timing}. The time difference between a proton and a pion with a momentum of 15~GeV/c after travelling one meter is 6~ps, well below the 30~ps resolution targets of the timing detectors. Therefore, the addition of these timing detectors will not affect the ability of the experiments to distinguish hadrons in these low-multiplicity decays and, although having the smallest effect on the peak position of the $W$ boson of those studied here, particle misidentification will remain an important factor to consider. 

\section{Partial reconstruction} 
\label{sec:pr}

\begin{figure}[t]
\centering                        
\includegraphics[width=6.5cm]{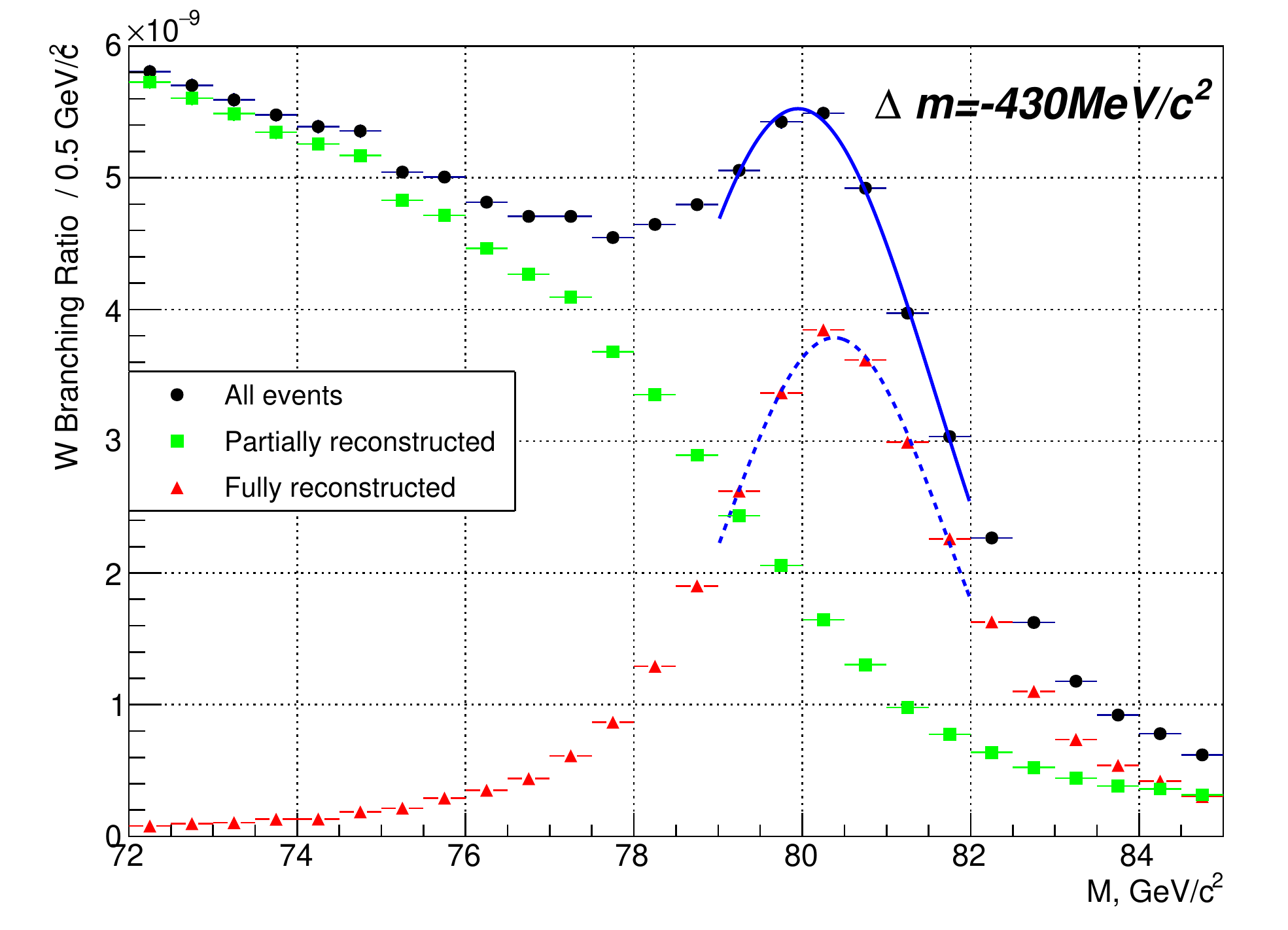}
\includegraphics[width=6.5cm]{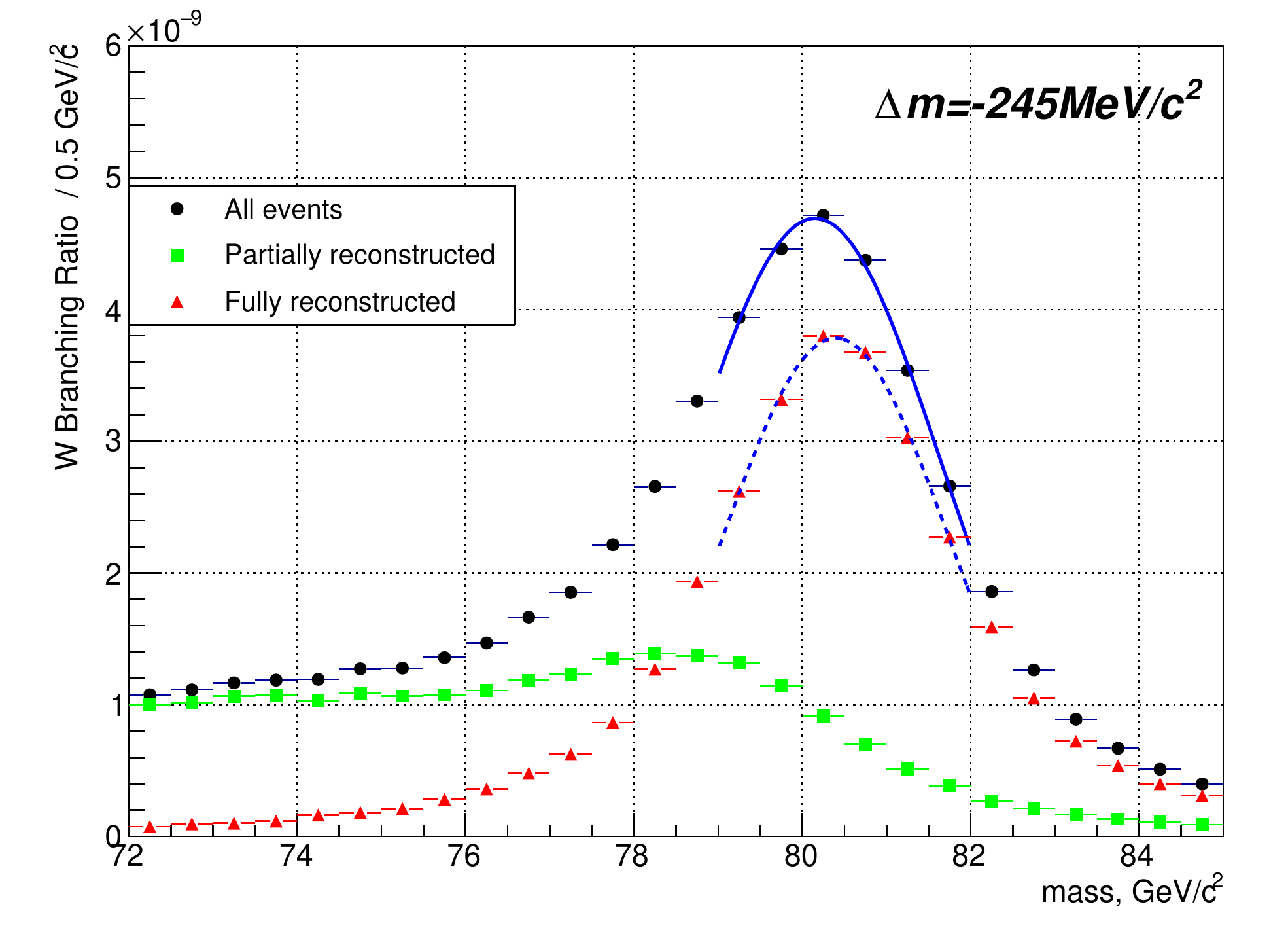}
\vskip -0.2cm                     
\caption[]{Left: The mass distribution predicted by Sherpa-ME-Lund for a $W$ boson decaying to a photon and three charged tracks. The fully-reconstructed events are in red, whilst those with additional neutral particles are shown in green, and the total in black. A detector resolution of 1\% is assumed. Right: the same distribution, except with a veto on $\pi^0$s with energy over 2~GeV applied. The mass region 79-82~\GeV~is fit with a Gaussian in order to extract the peak position and the fitted curve is shown in blue. \label{figs/Fig:partial} }
\end{figure}

Partial reconstruction refers to attempting to reconstruct a particle using a subset of the actual decay products. The example of reconstructing a $W^-$ boson using three charged tracks and a hard photon in the final state, $W^- \ra h^- h^+ h^-\gamma$, as predicted by Sherpa-ME-Lund, is shown in Fig.~\ref{figs/Fig:partial}, where $h^{\pm}$ refers to a charged hadron. 

The relatively large number of decays including additional neutral particles constitutes a major background that overlaps with the $W$ mass peak. A detector resolution of 1\% is assumed, but even with perfect resolution this overlap would be unavoidable due to the $W$ width. These partially reconstructed events could, in principle, be suppressed by vetoing on photon-like calorimeter energy deposits. ATLAS and CMS used 3~GeV and 1~GeV thresholds, respectively, for accepting energy clusters in their electromagnetic calorimeters late in the LHC Run 2~\cite{ATLAS19,CMSem}. A  $\pi^0$ with energy above 2~GeV will give rise to at least one photon with energy above 1~GeV, which suggests it might be possible that they could be efficiently vetoed in the CMS experiment. The effect of vetoing events with a $\pi^0$ above this energy is shown in Fig~\ref{figs/Fig:partial}~(b).  Although a large fraction of the partially reconstructed decays is removed, the veto is much less efficient close to the $W$ peak. The result of fitting Gaussian curves to the peak region is that the measured position is shifted by $-430\pm14$~\MeV, or $-245\pm12$~\MeV\ when the high-momentum $\pi^0$ veto is applied. The observed shift, when applying a veto, would depend upon the detailed calorimeter performance under intense pileup and, therefore, it would also have additional systematic errors.

The partially reconstructed event rate could perhaps be measured by reconstructing an additional $\pi^0$, although the mix of charged and neutral energy in the jet would degrade the resolution. Alternatively, the impact could be partially offset by fitting the low-mass sideband of the $W$ peak. However, in each case a reliable model is needed to relate the measured region to the unmeasurable, but most relevant, low momentum $\pi^0$s. 

Although it was not possible to simulate  $W^-\ra\pi^-\gamma$ or $W^+\ra D_s^+\gamma$, they  will also suffer from partial reconstruction due to the $W^-\ra\rho^-\gamma$ and $W^+\ra D_s^{+*}\gamma$ decays. This is particularly problematic in the case of the $D_s^{+*}$, where the dominant $D_s^{+*}\ra D_s^+\gamma$ decay produces a single photon of low energy.

\section{EW-QCD interference}
\label{sec:ewqcd}

Interference of the hadronic vector boson decay with the t-channel $\qq\ra\qq$ QCD background has been shown to introduce an apparent reduction in the mass of the peak~\cite{RANFT, BAUR, PUMPLIN, DING}. This comes from the change in sign of the Breit-Wigner resonance at the peak, overlaid on a continuous QCD background. The effects reported are mass shifts of hundreds of \MeV. 

\begin{figure}[htb]                  
\centering                        
\includegraphics[width=6.5cm]{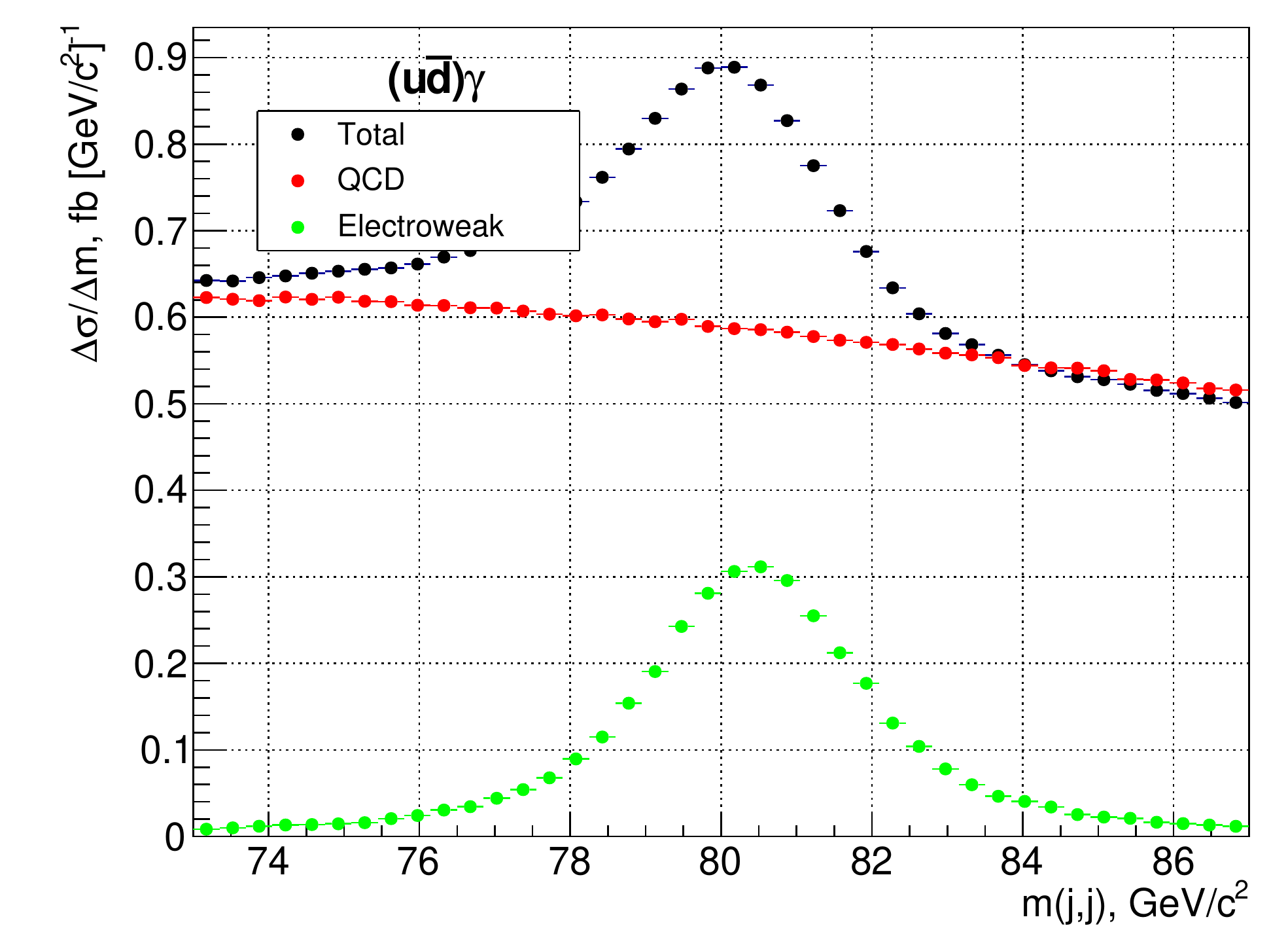}
\includegraphics[width=6.5cm]{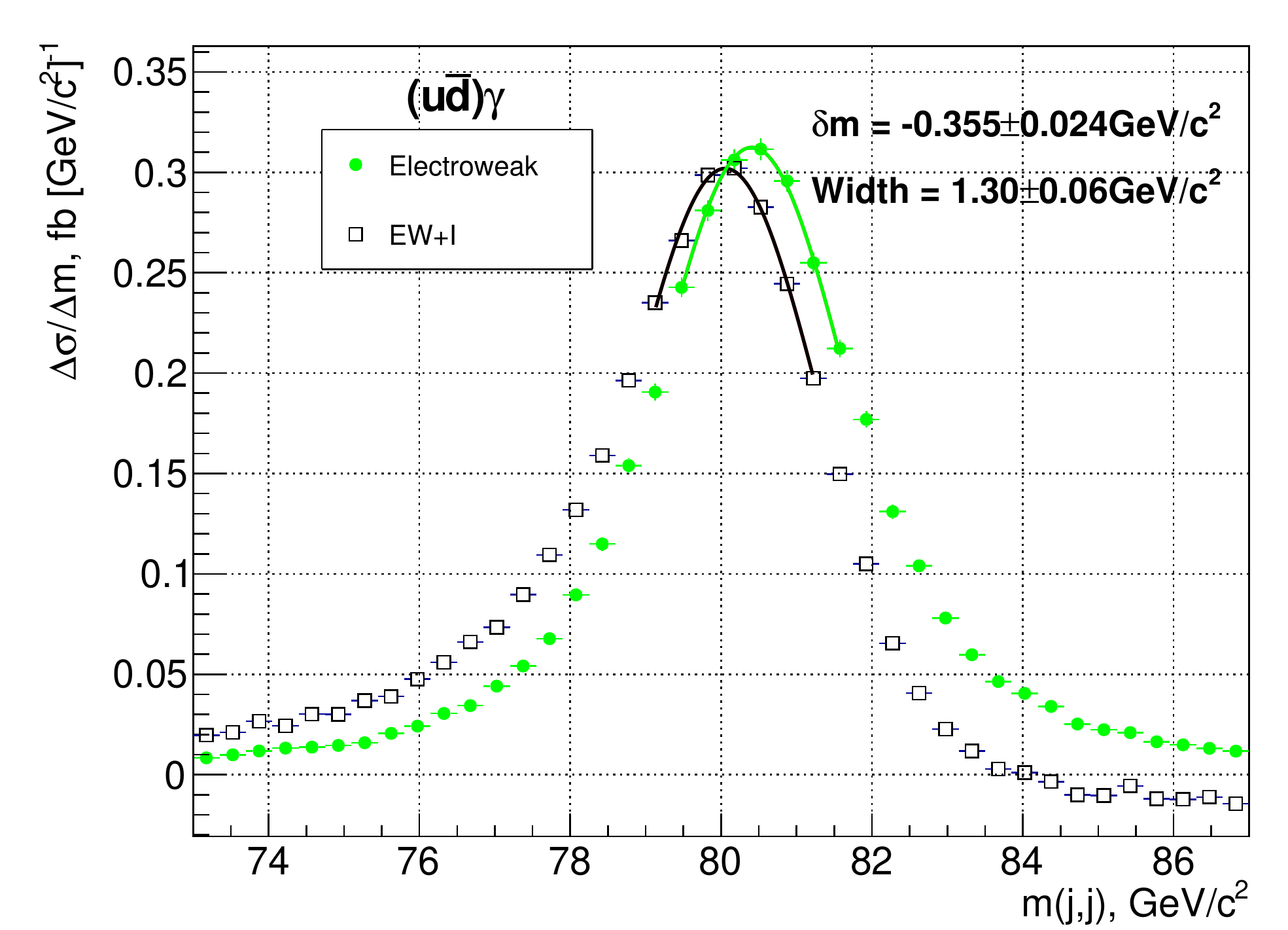}
\vskip -0.2cm                     
\caption[]{The reconstructed mass peak in the $\ud\gamma$ system, assuming a 1\% detector resolution and requiring the mass of the \ud\ system to be below 2~\GeV. Left: the basic distributions for the EW component (green), QCD component (red) and total (black). Right: A Gaussian fit to the EW (green) and total (black) terms to determine the mass shift. \label{figs/Fig:bias} }
\end{figure}

For radiative final states, the Leading Order (LO) Sherpa-ME-Lund setup is used, with the hadronization omitted to speed computation. For these interference studies, all final state particles were required to satisfy $|\eta|<2.5$, with the mass of the \ud\ system required to be below 2~\GeV and that of the $\ud\gamma$ system to be between 70 and 90~\GeV. Fig.~\ref{figs/Fig:bias} shows the mass predicted by this configuration for $W^+\ra\ud\gamma$ events from the EW process alone, from the QCD background process and from the coherent sum. The predicted shift in the mass peak is $-355\pm24$~\MeV, after including the detector resolution of 1\%. The mass shift in $W^-\ra\du\gamma$ is $-322\pm24$~\MeV, consistent with the $W^+$ but smaller, as expected with a slightly better signal-to-background ratio. 

The observed shifts are insensitive to the mass selection applied in the simulation of the di-quark system. They do, however, depend upon the fiducial region chosen, with the $|\Delta \eta|$ between the hadronic system and the photon very important. They need to be evaluated for each specific experimental configuration.  It must also be emphasised that these calculations are at LO only. Thus, they are merely indicative of the scale of the effect and an NLO evaluation would be highly desirable. The observed shifts would be much smaller for $W$ from top decay, since the intrinsic signal-to-background ratio is much better.

The second generation decays, $W^+\ra\cs\gamma$ and $W^-\ra\sc\gamma$, have smaller QCD contributions and hence reduced interference. The shifts in the peak positions are consistent with each other, $-96\pm14$~\MeV\ and $-90\pm48$~\MeV, respectively.  In addition, an exploration of interference in the fully-hadronic decay mode was conducted. The final state \us\sd\ was simulated, with all four
quarks having a momentum over 2 GeV/c, the \us\ and \sd\ systems each required to have a mass below 
6~\GeV~and the four other quark pairings having a mass above 10~\GeV. As before, all quarks were required to satisfy $|\eta|<2.5$. This channel was chosen since it arises naturally in  $W^+$ decay and the quarks are all distinct, which simplifies the simulated phase space definition. The EW signal cross-section, within this (arbitrary) mass cut, is similar to the radiative process, whereas the QCD component is approximately  one hundred times larger, although this can be varied by kinematic selections such as restricting $|\Delta\eta|$. The worse signal-to-background ratio makes the fully-hadronic channel much less promising than the radiative channels for LHC searches. A shift in the $W$ mass peak position in the fully hadronic channel of $-970\pm290$~\MeV\ was seen.

\section{Conclusions}
\label{sec:conc}

Radiative $W$ boson decays are predicted by PYTHIA8 and Sherpa for three-body and four-body final states, with branching ratios of a few in $10^{-8}$. This is three orders of magnitude below the current
sensitivity for the experimentally similar decay of $W^-\ra\pi^-\gamma$.  It is possible, however, that exclusive hadronic vector boson decays may be observed at the (HL-)LHC. Radiative decays, with a high-energy photon recoiling against a low-multiplicity hadronic system, seem to be the most promising channels. Although it might be technically feasible to trigger fully-hadronic $W$ boson decays, the poor intrinsic signal-to-background ratio means it is unlikely to be competitive with the radiative decay modes. 

In studying such states, three effects need to be understood and controlled: charged hadron misidentification, partial reconstruction and interference. Misidentification is the least significant of these effects; even so, it can produce shifts in the peak mass position of the order of 60~\MeV. The effect of partial reconstruction was found to reduce the peak position by approximately 400~\MeV\ for decays to $h^+ h^- h^+\gamma$. However, the parton shower simulations have little guidance from data and are close to their kinematic limit, so this prediction must be regarded as unreliable. Interference decreases the measured mass by 300 (100)~\MeV\ for \ud\ (\cs) final states, but it should be noted that these calculations were performed at LO, with no NLO estimate currently available. 

In summary, fully-reconstructed $W$ boson decays have significant hurdles that need to be well understood in order for a mass measurement in these channels to be feasible. The effects studied here  should also be considered in searches for such decays.

\end{document}